# Life and space dimensionality:
# A brief review of old and new entangled arguments

Francisco Caruso[1]
Centro Brasileiro de Pesquisas Físicas – CBPF
Rua Dr. Xavier Sigaud, 150 – Urca, Rio de Janeiro, RJ, 22290-180, Brazil

**Abstract:**

A general sketch on how the problem of space dimensionality depends on anthropic arguments is presented. Several examples of how life has been used to constraint space dimensionality (and vice-versa) are reviewed. In particular, the influences of three-dimensionality in the solar system stability and the origin of life on Earth are discussed. New constraints on space dimensionality and on its invariance in very large spatial and temporal scales are also stressed.

---

[1] Email: caruso@cbpf.br; francisco.caruso@gmail.com

**Preamble**

More and more researchers devote themselves to understand or at least shed some light onto two apparently uncorrelated scientific issues: the origin of life[1-6] and the threefold nature of physical space.[7-9] The first is *per se* a fascinating subject and probably one of the most difficult problems to be solved; the second is related to the dream of a Grand Unification in Physics. The issue of the origin of life has a continuously renewed interest due to improvements in the accuracy of measuring instruments. Another contribution comes from new data originated from space telescopes and satellites, as COBE, as will be stressed in this paper. The discovery of planets beyond our Solar system using data from the new ground and space-based telescopes, such as Kepler, is an additional source of interest. Furthermore, the interest on the quest of space dimensionality is now being renewed as the search for experimental evidences of extra dimensions is part of the contemporary tendency to investigate physics "beyond the Standard Model" in Collider Experiments.[10] A modern and comprehensive survey of dimensionality can be found elsewhere.[11]

From a modern perspective, the idea these two open problems can be somehow entangled can be traced back to Schrödinger's seminal ideas collected in his 1944 famous book *"What is life?"*.[12] There, it is claimed that life should be understood in terms of the new Quantum Physics, relating the stability of genes to the discontinuous (quantum) transitions of physical states. Therefore, it is expected a significant contribution of Physics for the understanding of life. In particular, the influence of the topology of physical space (like its dimensionality) on the kind of life we know should not be neglected.

In spite of this expectancy, there has been an enormous amount of speculation about the origin of life, with little heed to constraints that might be imposed by the physical settings.[3] This is particularly true concerning the influence of topological properties of space, like its dimensionality.

The main scope of this brief report is to summarize how physical and philosophical approaches to the problem of space dimensionality are related to life through a version of the Anthropic Principle. The paper also aims to discuss the temporal scale of each kind of restriction imposed onto space dimensionality. Finally, it attempts to argue that recent analysis of the microwave background radiation spectrum gives rise to the first constraint on dimensionality for a given temporal scale larger than that usually required for the existence of life.

**The historical roots**

That space is three dimensional seems to be so obvious to laymen and even to scientists that one can easily disregard it as a scientific problem. Indeed, in almost all physical and chemical theories developed along centuries, *dimensionality* – an essential topological feature of space – is merely assumed as a given truth, as an unquestionable matter of fact supported by visual, tactile and kinesthetic perception of space. However, in this review it will be shown that things can go in a different way, and that we can improve our comprehension about the possibility of having physical space dimensionality fluctuating or not in a large spatial and temporal scale. How this could affect the origin of life, and, vice-versa, how the conditions for the existence of life are used to constrain the dimensionality are issues that will also be addressed.

The way Aristotle and Kant caught a glimpse of the possibility of relating space dimensionality with the perceived World and life should be recognized and emphasized as two keystones.

It is well known that Aristotle did not develop a theory of *space*. He instead weaved a theory of place (*topos*), which is mainly discussed in his *Physics*. In a modern language, the Aristotelian *topos* has always been a bi-dimensional surface (the inner limit of a containing body) and space has been a kind of collection of all possible places. However, it is not in this book that he treated the question of dimensionality, which is considered, instead, in his cosmological and biological texts, namely *On the Heavens* (*De Caelo*) and *Movement of Animals*. In fact, in the Book I of *De Caelo* he says that "we cannot pass beyond body to a further kind, as we passed from length to surface, and to surface to body'',[13] since he admitted body alone to be determined by the three dimensions. However, it was only in the *Movement of Animals* that the Stagirite tried to develop a theory of dimensions based on the study of movement. Since he had a hierarchical conception of space (much different from what we now call Euclidean space) he was led to consider in his biological text the existence of *six* and not *three* dimensions (up-down; forward-backward; left-right). Thus, he speculates that those dimensions, which for him were related to the soul (*psyche*), somehow *define* the nature of living beings.[14] To the best of our knowledge, this is the first attempt to relate space dimensionality to life.

Kant, in his first writing,[15] unsuccessfully tried to demonstrate[9,16,17] that the threefold nature of space is a consequence of the Newton's inverse square law of Gravitation. He was worried about the question of how to express the interaction of physical substances in universal terms of cause and effect, and in what way *matter* (the substance) is able to alter the state of the *soul* by means of the force it possesses in its motion. In Kant's opinion it is through these forces that connection among bodies can be established, from which the *order* necessary for the existence of *space* is achieved. This was the first step in the direction of a scientific explanation of dimensionality. Even though it has been shown [17] that Kant did not actually succeed in proving this conjecture – indeed, he just concluded that there should be a relationship between this law and *extension* –, his contribution has the very merit of suggesting that the problem of dimensionality can also be treated in the framework of Physics and does not belong exclusively to the domain of Mathematics, neither to that of pure philosophical speculation.

**Whitrow's biotopological justification of three-dimensionality**

In 1955, Whitrow asseverates that for trying "*to isolate three-dimensional space as the only possibility for the world in which we find ourselves, we must now invoke some argument for showing why the number of dimensions cannot be less than three*".[18] To do this, he adapted the well-known topological result from knot theory, that we cannot make a knot in an even-dimensional space, to the necessity of higher forms of animal life to have brains in which electrical pulse information carried on by nerves could not interfere destructively, which excludes a twofold and other even-fold spaces. This argument automatically constrains space to have an odd dimensionality $\geq 3$. Recognizing that the problem of space dimensionality was not yet solved – which is still true in our opinion – Whitrow wrote in the conclusion of his paper:

> *"Despite various recent attempts to show that [space dimensionality] is either a necessary attribute of our conception of physical space or is partly conventional and partly contingent, the problem cannot be considered as finally solved. A new attempt to throw light on the question indicates that this fundamental topological property of the world may possibly be regarded as partly contingent and partly necessary, since it could be* inferred *as the unique natural concomitant of certain other contingent characteristics associated with the evolution of the higher forms of terrestrial life, in particular of Man,* the formulator of the problem.*"*[18]

Following a different approach, based on the *stability of atoms in high dimensional spaces*[19-22] and on the Uncertainty Principle, Barrow & Tipler stressed that

> *"(...) it has been claimed that if we assume the structure of the laws of Physics to be independent of the dimension, stable atoms, chemistry and life can only exist in N<4 dimensions."*[23]

And therefore they conclude, perhaps inspired on the aforementioned Whitrow's ideas, that '*the dimensionality of the Universe is a reason for the existence of chemistry and therefore, most probably, for chemists also*'.[24]

So, chemists should be proud, first of all just because they exist and also because their existence should, somehow, be related to the comprehension of space dimensionality. This is not, however, a completely original idea; it is related to the so called "Anthropic Principle".

**The setup of an anthropic framework for the modern discussion of space dimensionality**

To the best of our knowledge, the expression "Anthropic Principle" was coined, in 1973, by the astrophysicist Brandon Carter as a sort of reaction to the tremendous impact of Copernican Revolution on both Science and Society that took Men out of the center of the Universe.[25] As Carter himself stressed, '*although our situation is not necessarily central, it is inevitably privileged to some extent*'.

Nowadays, this expression hides many different meanings. What is now known as the "Weak Anthropic Principle" has its origins in an earlier Dicke's idea,[26] which was reformulated by Carr & Rees.[27] The new focus essentially tells us the following: observed values of physical quantities are not arbitrary but restrict to be compatible with the sustained evolution of life so far spatiality is concerned, and temporally consistent with biological and cosmological evolution of living beings and of their niches. There is also the "Strong Anthropic Principle" due to Carter,[28] which assumes that the Universe necessarily should contain life, and the "Participative Anthropic Principle" advocated by Wheeler who sustain that Observers are necessary for the existence of the Universe. Such argument is a sort of consequence of the measurement problem in Quantum Mechanics.[29] More details can be found in Bettini's paper.[30]

In any case, of relevance to the present review is the very fact that anthropic arguments have been proposed, independently, by philosophers and scientists to explain why we perceive a three-dimensional Universe.[31] We could even say more: it seems unavoidable to make use – implicitly or explicitly – of some anthropic argument when we try to justify and to understand three-dimensionality. Some of these proposals will be briefly reviewed in this Section.

One can quote the work of William Paley,[32] at the beginning of 19th century, as an important attempt to shed light on the space dimensionality problem from Anthropic arguments. In his work, Paley analyzes the consequences of changes in the form of Newton's gravitational law and of the stability of the solar system on human existence. Starting from a teleological thesis, his speculations take into account a number of mathematical arguments for an anthropocentric design of the World, which rest all upon the stability of the planetary orbits in our solar system and on a Newtonian mechanical *Weltanschauung*, as should be expected at that time.

Actually, both are typical ingredients of an anthropic constraint imposed on dimensionality. In spite of the fact that this kind of approach strongly reflects the recognition of our ignorance be complete and assumes a 'Principle of Similarity' – using the expression adopted elsewhere,[18] namely that alternative physical laws should mirror their actual form in three dimensions as closely as possible. The form of the differential equation describing a particular physical phenomenon is supposed to be equally valid in other dimensions. The structure of the equation is maintained and only the number of dimensions is changed. However, it is clear that we cannot demonstrate, for example, that Poisson's equation, which is satisfied by a Newtonian gravitational potential, should still have second order derivatives in other dimensionality. Such choice can only be justified by some kind of anthropic principle. It seems a very hard task to avoid this hypothesis as long as dimensionality is to be understood in the realm of Physics or Chemistry, as pointed out in the next Section.

**The stability of atoms**

In the twentieth century, the idea of how space dimensionality follows from the stability of planetary orbits in the solar system was revisited in two Ehrenfest's seminal papers,[19,20] where several physical phenomena were discussed, trying to disclose any qualitative difference between three and other n-dimensional spaces. The existence of stable atoms, molecules and planetary orbits is just one point. These space dimensionality depending aspects, which distinguish Physics from one dimensionality to another, are called by him "singular aspects" and his works were aimed at stressing them. A crucial assumption is built into Ehrenfest's ideas, namely that it is possible to make the formal extension from three to n-dimensional space for a certain law of Physics and, then, one should find one or more principles that, in conjunction with this law, can be used to single out the proper dimensionality of space. The generality of this approach was noted by Tangherlini[21] who proposed that for the Newton-Kepler problem, generalized to n-dimensional space, the principle for determining the spatial dimensionality could be summarized in the postulate that *there should be stable bound states orbits* – or "states" – for the equation of motion governing the interaction of bodies, treated as material points. This will be generically called, from now on, the *stability postulate*. Later, Tangherlini showed that the essential results of the Ehrenfest-Whitrow investigation are unchanged when Newton's gravitational theory is replaced by General Relativity.[22] In this way he attributes a new scientific meaning to Paley's conjecture. Application of this same idea to the stability of hydrogen atom, described by a generalized Schrödinger equation, leads to the same kind of constraint in a very huge and different spatial scale, *i.e.*, from planetary to atomic scales. Considering the stability of both non-relativistic and relativistic hydrogen atoms the reader should be aware of new results qualitatively different from Tangherlini's.[33-34]

This briefly reviews how the *stability postulate*, in which an anthropic argument is implicit, is used to cast some light on the problem of spatial dimensions. However, one should point out that some epistemological and methodological aspects of this general approach based on the stability postulates were criticized twenty years ago.[9]

**Some old and new arguments**

Each one of the constraints on space dimensionality derived from the *stability postulate* is valid for a particular spatial and temporal scale. This means that each argument based on the origin or on the existence of life tells us that space dimensionality should be three at some scale. In fact, looking at the past, one can estimate, case by case, the interval of time T and the spatial region characterized by a length L for which one can surely say that the dimensionality compatible with life is three.

Starting from the already quoted results, Ehrenfest's stability argument is typically valid for distances of the order of the solar system and in a time scale large enough to make the evolution of life possible on Earth, as mentioned by Whitrow.[18] However, Ehrenfest's argument about this subject could be questioned by stressing that it is not sufficient that the intensity of solar radiation on Earth's surface should not have fluctuated greatly for life still exist on Earth; actually, the fact that the Sun's radiation spectrum did not fluctuate very much should also be required.[9] This argument can be corroborated if we remember that atomic spectra are observed in galaxies with a redshift (z) corresponding to c. 600 million years from the Big Bang.

On the other side, Tangherlini's work about the stability of hydrogen atoms is often invoked to suggest the validity of Chemistry in the same time scale as a necessary, although not sufficient, condition – at least Chemical Thermodynamics of irreversible processes should be also valid. Thus, '*the presence of atomic spectra in remote stars may also indicate[s] that space has had the same dimensionality at cosmic scale*'.[35] The existence of such an astrophysical constraint on space dimensionality is treated elsewhere.[36]

Another class of arguments is also related to the general idea that, among a large number of possible universes, the actual Universe is the one which contains intelligent life or, at least, had some form of life in a very long time scale. We have previously mentioned what Withrow, Barrow and Tipler said about human life and how it imposes some constraints on the number of dimensions. Inevitably this query refers us to Biochemistry. Barrow and Tipler's 1986 book has a nice chapter on this subject where several relevant topics are discussed in details and so they will not be treated here. Among them we can quote just the unique properties of carbon, hydrogen, oxygen and nitrogen, or whether or not it is possible to base life on elements or substances other than these ones and, finally, that these unique properties are probably necessary to guarantee the ecological stability required by highly-evolved life, although not sufficient.

Our aim here is to introduce a new argument in favor of a stable scenario for space dimensionality considering a temporal scale longer than that required for the existence of human or another kind of highly-evolved life on Earth. The usually accepted scales is that the *homo erectus* appeared 2 million years ago, while the first *skeletons and easily recognizable fossils* range are date 600 million years ago. This temporal scale can be enlarged taking into account the detection of Policyclic Aromatic Hydrocarbons (PAHs) from objects with $z \simeq 3.2$, which corresponds to a lookback time of 11.8 billion years. In *summa*, we can anticipate that this new argument is related to the tetrahedral structure of carbon as will now be shown.

Let us consider the famous experimental result published by Miller,[37] in 1953 – see also the 1959 papers by Harald C. Urey and Stanley Miller.[38] In a certain sense, we can understand these works as natural extensions of Urey's concern[39] about the origin of the solar system and the chemical events associated with this process. They showed to be possible, by means of an electrical discharge, to transform an admixture of gases consisting of methane, water, ammonia and hydrogen – believed to be, at that time, the composition of early Earth atmosphere – into a relatively small number of biochemically relevant compounds, among them hydroxyl acids, urea and some amino acids essential to life, as glycine and alanine. Although it is not a *proof*, this historical result is widely considered as a strong evidence for the creation of life in a kind of primitive Earth atmosphere composed of the four substances just mentioned, quite different from that of the present days. Accepting this might suggests that, in certain sense, methane, which has the simpler formula among the organic compound ($CH_4$), is somehow related to the origin of amino acids that could build up primitive life.[40] In addition, it is important to stress that a strong assumption is implicit in this reasoning, namely that the atomic structure and chemical properties of the elements have not changed in time (*principle of similarity*). Actually, this assumption may be supported by looking to the star spectroscopy firstly discovered by von Fraunhofer and explained by Bunsen and Kirchhoff as an experimental endorsement to the hypothesis that each different atom is strictly the same in any part of the Universe. In any case, such kind of hypothesis together with the conjecture that all physical laws did not varied during the Universe evolution are both essential ideas underling the *Cosmological Principle*, which obviously depends on an observer, otherwise it would be very difficult to describe this evolution in terms of Physics.

More recent studies have shown that some amino acids and sugars could have come to Earth from interstellar medium, as reviewed by Marshall,[41] and reported by others.[42-43] For example, some meteorites that collided with Earth after a long journey from the remotest corner of Solar system and beyond contained amino acids. This is the case of the meteorite that struck Murchison, Australia, in 1969. Indeed, trace amounts of glycine, alanine, glutamic acid, valine and proline were detected quite immediately by a conventional ion exchange chromatography.[44] On the other hand, it has been well established that amino acids may be found in all most carbon-rich meteorites.[45] This is a relevant point for the subject of this paper since

> *"amino acids in meteorites have naturally attracted attention because of the central role that such acids play in terrestrial biochemistry, and the possibility that both meteoritic and prebiotic terrestrial amino acids shared a similar origin"*.[46]

In 2002, another group reported a laboratory demonstration that glycine, alanine and serine naturally form from ultraviolet photolysis of the analogues of icy interstellar grains. This ice is primarily composed of amorphous $H_2O$, but usually also contains a variety of other simple molecules, such as $CO_2$, $CO$, $CH_3OH$ and $NH_3$.[47] The idea that spontaneous generation of amino acids in the interstellar medium is possible also seems to be supported by another independent result.[48]

Therefore, we have seen at least four different results which strongly suggest a mechanism of amino acid creation in the interstellar medium. Putting those evidences together with the result of Urey-Miller's experiment, we realize that, in general, it is not the methane that plays a crucial role in amino acids synthesis. The most common simple chemical fact underlying all these experimental results is the presence of *carbon* on a previous amino acid free medium, which can be interstellar or from the primeval Earth atmosphere.

On the other side, based on X-ray spectroscopy and on the empirical fact that an isomer of methane has never been found, the tetrahedral structure of carbon was established. In other words, Nature seems to have chosen just one spatial disposal for methane atoms and also for all compounds of the type $CH_3Y$ e $CH_2YZ$, with Y and Z being any group of atoms, confirming the intuition of van't Hoff.[49] This rules out any flat configuration for the simplest organic compound and other carbon-made molecules and requires, obviously, that the space in which it exists should be (at least) three dimensional. Thus, just the amino acids synthesis – no matter how sophisticated mechanisms are necessary to build up more complex molecules in the chain of life – , like we understand it today, presupposes carbon to be bound in tetrahedral structures and requires (at least) three-dimensionality of space. This puts the temporal limit of such a constraint on three-dimensionality at least in the order of something like 3,500 million of years ago (expected temporal scale for the origin of life in the form of thermofiles), which, *grosso modo*, is larger than what is expected from Whitrow's bio-topological argument (admitted here as requiring the existence of higher forms of animal life) by a factor of $10^3$.

Therefore, believing on Urey-Miller's experiment and on the laboratory synthesis of some amino acids from a medium similar to the interstellar one as a clue for the origin of amino acids essential to life, associated to an atmosphere with a significant amount of carbon, implicitly assumes that *three* is the minimum space dimensionality required by the tetrahedral carbon structure and for life to be developed. Putting this together with what was previously said about the spectra of remote stars, a scenario where space dimensionality should be at least three for very large temporal and spatial scales seems plausible; much greater than that required by human life on Earth. Despite its speculative nature, this is a new constraint imposed not only on the number of dimensions but also on its stability throughout a very large spatial and temporal scale, obtained from a sort of modified strong Anthropic principle, namely from the assumption that the early Universe should necessarily contain amino acids.

We can still compare this result to a theoretical analysis of the cosmic microwave background spectrum measured by FIRAS (Far InfraRed Absolute Spectrophotometer) on COBE satellite,[50] which established a very tiny limit for the possible deviation of space dimensionality from 3, namely something of the order of $10^{-5}$. Such deviation can be thought as a limit for fractal dimensions following the ideas of Mandelbrot[51] or simply viewed as the fluctuation of a measurement.

More precisely, in this paper space dimensionality D is supposed to be $D=3+\varepsilon$, and data fitting give us $\varepsilon = - (0.957 \pm 0.006) \times 10^{-5}$. The value of $|\varepsilon|$ can also be interpreted as an upper limit for how much space dimensionality could have deviate from three. This means that any deviation of space dimensionality from the well accepted value 3 must be very small and took place in a very large spatial and temporal scale, comparable to that of the decoupling era. This is indeed a constraint valid at all scales comprised between the decoupling era and today. However, one should stress that this is strictly true in the classical scenario of General Relativity, because topological changes (including space dimensionality) are forbidden in the sense that their presence would necessarily imply the appearance of either singularities or closed time-like curves. This result is known as Geroch Theorem.[52] An alternative where the value of D could change in time and space may result if quantum fluctuations are considered in a Quantum Theory of Gravitation.

Concerning the general consequences of such deviations, we agree with the concluding remarks of the Zeilinger & Svozil's paper:[53]

> "*It is certainly a challenge for future research to investigate whether or not the deviation of the dimension of space-time from four can be made more statistically significant than the present work suggests. Furthermore, the question of possible evidence for such a small deviation in other areas of Physics deserves attention*".

Meanwhile, if we assume time to be one-dimensional, as usually done, the main result of Caruso & Oguri[50] can be seen as an answer to both challenges. Finally, since the small deviation from three-dimensionality of space obtained here is extracted directly from the CMBR data, it actually suggests, as already said, that space dimensionality did not vary significantly in a huge temporal scale, once this background radiation is expected to be related to the Big Bang. This time scale can be safely put on the later epoch where the universe was about $3 \times 10^5$ yr old ($z \simeq 10^3$). In addition, due to its isotropy, this is the *only* experimental situation that, being confronted with *any* other local experiment on Earth, aimed to measure the number of space dimensions, drives to the conclusion that dimensionality did not vary also in a very large spatial scale (a cosmological scale).

Everything we have discussed in this paper leads us to build a coherent scenario where space dimensionality has been three (or a very close number) since the decoupling era that happened after the Big Bang, and this is exactly the value necessary to the primeval synthesis of amino acids as well as for the origin of life on Earth. These evidences just corroborate the theoretical idea that space dimensionality is a topological invariant that could not be changed in the framework of General Relativity.

**Concluding remarks**

There are still a lot of scientific and philosophical work to be done concerning both the problem of space dimensionality and that of the origin of life. From the point of view of Physics, two of them can be emphasized here. Studies of how topological aspects of physical spaces could be determinant to the forms of life known until now or to its origin will be welcome. Secondly, we should be aware of the biased incompleteness in the majority of approaches to this problem, once they consider physical events taking time to be one dimensional and the equations which describe such events are just postulated to be equally valid for any dimensionality. Actually, it is well known that in the framework of any relativistic theory, space dimensions are not independent from time dimensions. This means that the usual (classical) consideration that time can be *a priori* fixed to be one-dimensional should be abandoned. Finally, a deeper comprehension on the problem of space-time dimensionality is still to be reached. In particular, if it could be possible to go on discussing this problem without taking into account any kind of anthropic argument, as some stage of a particular reasoning is an open question yet. Therefore, we conclude saying that there are still good questions without good answers.

**Acknowledgments**

The author would like to thank Oscar Matsuura, Gilvan Alves, Hélio da Motta, Mauro Velho de Castro Faria and Maria Lucia Bianconi for fruitful discussions and for critical comments.